\documentclass[a4paper]{article}

\usepackage[english]{babel}
\usepackage[utf8x]{inputenc}
\usepackage[T1]{fontenc}

\usepackage[a4paper,top=3cm,bottom=2cm,left=3cm,right=3cm,marginparwidth=1.75cm]{geometry}

\usepackage{amsmath}
\usepackage{graphicx}
\usepackage[colorinlistoftodos]{todonotes}
\usepackage[colorlinks=true, allcolors=blue]{hyperref}
\date{}
\title{A scenario for the Galactic cosmic rays \\ between the knee and the second-knee}
\author{Silvia Mollerach and Esteban Roulet\\
Centro At\'omico Bariloche, Comisi\'on Nacional de Energ\'\i a At\'omica\\
Consejo Nacional de Investigaciones Cient\'\i ficas y T\'ecnicas (CONICET)\\ 
Av. Bustillo 9500, R8402AGP, Bariloche, Argentina}

\begin{document}
\maketitle

\begin{abstract}
We perform a fit to measurements of the cosmic ray spectrum in the energy range between $10^{15}$~eV and  $10^{18}$~eV using data from the TALE, Tunka and Auger experiments. We also fit  the data on the depth of shower maximum,  $X_{\rm max}$, from Tunka and Yakutsk or from  Auger to constrain the cosmic ray composition.
 We consider a Galactic component that is a mixture of five representative nuclear species (H, He, N, Si and Fe), for which we adopt rigidity dependent broken power-law spectra, and we allow for an extragalactic component which becomes strongly suppressed for decreasing energies. The relative abundances of the Galactic components at $10^{15}$~eV are taken to be comparable to those determined by direct measurements at $10^{13}$~eV. The main features of the spectrum and of the composition are reproduced in these scenarios. The spectral knee results from the break of the H spectrum at $E_{\rm k}\simeq 3\times 10^{15}$~eV, although it is broadened by the comparable contribution from He which has a break at about $6\times 10^{15}$~eV. The low-energy ankle at  $E_{\rm la}\simeq 2\times 10^{16}$~eV coincides with the strong suppression of the H and He Galactic components and the increasing relative contribution  of the heavier ones, but the observed hardening of the spectrum at this energy turns out to result from the growing contribution of the extragalactic component. The second-knee at  $E_{\rm sk}\simeq 26 E_{\rm k}\simeq 8\times 10^{16}$~eV is associated with the steepening of the Galactic Fe component. The transition to the regime in which the total cosmic ray flux is dominated by the extragalactic component takes place at an energy of about $10^{17}$~eV. The parameters of the fit depend on the specific  $X_{\rm max}$ dataset that is considered, with the Yakutsk and Tunka data leading to a suppression of the light components being steeper beyond the knee so as to allow a faster growth of the average mass around the low-energy ankle. The results also depend  on the hadronic model that is used to interpret  $X_{\rm max}$ measurements and we compare the parameters obtained with Sibyll 2.3, EPOS-LHC and QGSJET II-04.
 The impact of   the possible existence of a maximum rigidity cutoff in the Galactic components is also discussed.
\end{abstract}

\section{Introduction}

The origin and nature of the cosmic rays (CRs) is still uncertain, even if more than a century has passed since their discovery. Cosmic rays are observed with energies extending from below $10^8$~eV up to more than $10^{20}$~eV and they consist mainly of different atomic nuclei having a composition (i.e. a relative abundance between different nuclei) that depends on the energy considered. Many different types of detectors are used to study them, from those put in balloons or satellites that detect directly the primary particles, the telescopes that observe the Cherenkov light emitted by the relativistic charged particles in the air showers produced when the primaries interact in the atmosphere or the fluorescence light produced by the nitrogen air molecules  that get excited by the passage of the showers, or alternatively arrays of detectors that measure the secondary particles reaching ground level. The direct detection of the primaries is feasible up to energies of about $\sim 10^{14}$~eV, above which the fluxes become too low. On the other hand, the observation of  the air showers with the indirect techniques requires very energetic primaries and can hence be used to explore the higher energy range above about $10^{15}$~eV (for a recent review see \cite{mo18}).   

Most of the CRs observed up to at least $\sim 10^{17}$~eV are believed to be produced in the Galaxy, possibly in association with supernova explosions and accelerated for instance in the shocks that are present in the supernova remnants or by electrostatic acceleration in the pulsars, while at energies higher than about $\sim 10^{19}$~eV they are believed to be of extragalactic origin, produced in  more violent sources such as active galactic nuclei or gamma ray bursts. The exact energy of the transition between the Galactic and extragalactic components is still a matter of debate.

The measured total CR flux follows approximately a power-law, d$\Phi/{\rm d}E\propto E^{-\gamma}$, with the spectral index being $\gamma\simeq 3$ although some significant changes of its value are observed at different energies. Below about 20~GeV/n (GeV per nucleon), the CR fluxes get strongly suppressed by the effects of the solar wind, which drags them away from the solar system and eventually prevents them from reaching the Earth\footnote{At these energies also some energetic solar particles may be produced in association with flares and coronal mass ejections in the Sun.}. Above these energies and up to about few $10^{15}$~eV, the overall spectrum has $\gamma\simeq 2.7$ although some differences are observed between the spectra of different components.    Note that both the CR acceleration at the sources as well as the diffusive  propagation in the Galaxy are due to  electromagnetic effects and hence depend in general on the rigidities $R$ of the particles, where $R=p/eZ\simeq E/eZ$, with $eZ$ being the charge of the particle with atomic number $Z$. However, for energies per nucleon smaller than about a TeV, the spallation processes due to the interactions of the CR nuclei with the interstellar gas are relevant, leading to a hardening of the spectra of the nuclei which suffer significant spallation, such as Fe nuclei, because their residence time inside the Galactic disc (and hence their spallation rate) decreases for increasing energies. For the same reason the spectra of the spallation products should become steeper, and this is quite apparent in the spectra of nuclei such as Li, Be or B \cite{amsb2c} which at low energies get their main contribution actually from the spallation of heavier nuclei such as C, N and O. Also a slight hardening of the spectra of the different components seems to be present at rigidities of about 300~GV  \cite{amsh, amshe, cream3}, although it is not clear if this fact is related to changes in the acceleration properties, to changes in the diffusive propagation or is due to other causes. A very characteristic feature which was observed long ago \cite{ku59}, known as the knee, is the steepening of the total spectrum taking place at an energy of few PeV, at which the spectral slope changes from $\gamma\simeq 2.7$ to a value of about three.  More recently, a hardening from $\gamma\simeq 3.1$ to $\gamma\simeq 2.9$ was observed at about 20~PeV \cite{leankle,tu133} and it is known as the low-energy ankle. There is a second-knee at about $10^{17}$~eV at which the spectrum steepens to $\gamma\simeq 3.2$ (see \cite{be07} for a review) to then harden again at the ankle \cite{li63}, at about 5~EeV, above which $\gamma\simeq 2.6$. Finally, the spectrum gets strongly suppressed above 40~EeV \cite{hiresgzk,augergzk,taspect}, tending towards an effective power-law index $\gamma\simeq 5$. Whether the spectrum has a final cutoff at energies beyond few hundred EeV or if instead it eventually recovers is still not known.

The knee feature has been interpreted as the steepening of the light Galactic component \cite{an05},
as would be the case if the spectra from the different mass components have a rigidity dependent break. This behavior could be due to a less efficient acceleration at the sources or to a more efficient escape from the Galaxy (or both). 
It is then expected that the spectra of the heavier nuclei of charge $Z$ should steepen at energies $ZE_{\rm k}$, where $E_{\rm k}\simeq 3$ to 5~PeV is the knee energy. This behavior, known as the Peters cycles \cite{pe61}, should lead to an increasingly heavier average composition for energies above $E_{\rm k}$. The low-energy ankle, at $E_{\rm la}\simeq 20$~PeV, may then coincide with  the transition between the steeply falling H and He spectra towards an increasingly  heavier composition. The steepening of the Galactic Fe component, which is the heaviest nucleus having a sizable abundance, is expected to be responsible for the second-knee at an energy $E_{\rm sk}\simeq 26E_{\rm k}\simeq 100$~PeV \cite{ca02}. A hardening in the light component taking place at $\sim 100$~PeV has also been reported by the KASCADE-Grande Collaboration \cite{ap13}, and it could be associated to a growing extragalactic contribution.

In this work we consider in detail the energy range between 1~PeV and 1~EeV, which contains the knee, the low-energy ankle and the second-knee. We search to reproduce these different features, as well as the observed composition trends, based on a model for the Galactic cosmic rays consisting of rigidity dependent broken power-law spectra, adding also a phenomenological parametrization of the extragalactic component that becomes relatively important above about 100~PeV. We consider the spectrum measurements obtained  by the Telescope Array Low-Energy extension (TALE)  in most of this energy range \cite{tale}. This experiment detects the Cherenkov light emitted by  air showers with energies between 2~PeV and up to about 100~PeV while  at higher energies it observes their fluorescence emission.  We also include the results of the Tunka-25 Cherenkov detector that extends the spectrum down to 1~PeV \cite{tu25} and at the highest energies, between 0.3 and 1~EeV, we include the results  from the Infill sub-array of surface detectors of the Pierre  Auger Observatory \cite{augersp}.\footnote{We have actually not included in this work, neither in the fits nor in the plots,  the two highest  energy bins of Tunka, since they contain just one event each, as well as the two lowest energy bins of the TALE spectrum because they are affected by large systematic effects and they lead to large contributions to the $\chi^2$.}  We also include measurements of the depth of shower maximum, $X_{\rm max}$, corresponding to the depth in the atmosphere along the shower direction, measured in g/cm$^2$, at which the electromagnetic component of the air shower reaches its maximum development. Since in a first approximation a shower produced by a nucleus of energy $E$ and mass number $A$ can be considered as the superposition of $A$ showers of energy $E/A$ (which being less energetic are also less penetrating), the average values of $X_{\rm max}$ at a given energy provide an indication of the average CR mass composition. We  use measurements of $\langle X_{\rm max}\rangle$ obtained with telescopes in different experiments, in particular those of Yakutsk \cite{yak} and  Tunka-133 \cite{tu133} or Auger \cite{augerxm}.
We normalize the relative abundances  at PeV energies with representative values comparable to those determined by direct measurements at $\sim 10$~TeV and, through a global fit, we determine the spectral features of the different Galactic components as well as the properties of the transition towards a dominant extragalactic component.\footnote{Several other experimental results exist in this energy range, but we do not attempt to perform a global fit to all existing data. In particular, we do not use composition measurements from surface particle arrays such as KASCADE, since the analysis of the hadronic model dependence, that would impact on the hadronic, electromagnetic and muonic components measured at ground, would be quite involved in those cases.} The interpretation of the $X_{\rm max}$ measurements depends on the model adopted to describe the hadronic interactions in the shower. We hence obtain results for three recent tunes of hadronic models:  Sibyll 2.3 \cite{sibyll}, EPOS-LHC \cite{epos} and QGSJet~II-04 \cite{qgsjet}.

 We note that related studies have been performed in terms of rigidity dependent galactic components and an extragalactic flux to account for Tunka data \cite{tu133} or for TALE spectrum data alone \cite{ab18}. Our more comprehensive analysis combines different datasets to extend the energy range covered, studies the implications from different composition results as well as their hadronic model dependencies, obtaining the parameters of the rigidity dependent scenarios from a global fit to the experimental results. 
A global spline fit to many spectrum and composition data has also been presented recently \cite{de18}, providing a parameterization of the different mass components in a wide energy range. See also \cite{ho04} for a review of earlier models to explain the knee feature.

\section{Spectrum of the Galactic and extragalactic components}

The Galactic CRs include almost all known nuclear elements, with a composition having some similarities with the abundances found in the solar system although they also present some clear differences (such as the abundances of the spallation products mentioned before or those that may be associated to the contributions to the CR fluxes coming from different types of supernovae, etc.). The most abundant elements are H, $^4$He, light nuclei such as $^{12}$C, $^{14}$N and $^{16}$O, intermediate mass nuclei such as $^{20}$Ne, $^{24}$Mg, $^{28}$Si or $^{32}$S or heavier nuclei such as $^{36}$Ar, $^{40}$Ca and specially $^{56}$Fe. We will hence adopt for simplicity five representative mass components, $A={\rm H}$, He, N, Si and Fe, to describe the Galactic CR fluxes. We model their spectra at energies beyond one PeV with rigidity-dependent broken power-laws such that
\begin{equation}
\frac{{\rm d}\Phi_{\rm G}}{{\rm d}E}=\sum_A\frac{{\rm d}\Phi^A_{\rm G}}{{\rm d}E}=\phi_{\rm G}\sum_A\ f_A \left(\frac{E}{\rm EeV}\right)^{-\gamma_1}\left[1+\left(\frac{E}{ZE_{\rm k}}\right)^{\Delta\gamma/w}\right]^{-w}.
\label{galflux}
\end{equation}
In this expression one has that for $E\ll ZE_{\rm k}$ the spectra of the components with different mass numbers $A$ scale as $E^{-\gamma_1}$. The factors $f_A$ represent the fractional contribution of the different elements at a given energy $E\ll E_{\rm k}$, and  are constrained by the condition $\sum f_A=1$.\footnote{Note that for instance $f_{\rm N}$ accounts for the CNO group, even thought the N abundance is much smaller than the C and O ones.} For reference, at energies of about 10~TeV the H and He contributions are similar and they amount to about 70\% of the CR flux, while the other three mass groups have comparable fractions, with the CNO abundances being about 50\% larger than those of the $Z=9$--16 group \cite{cream3,atic2,cream2,nucleon}. On the other hand, there are no indications of significant changes in the relative abundances taking place between 10~TeV and 1~PeV. We will hence adopt for definiteness the values  $f_{\rm H}=f_{\rm He}=0.35$, $f_{\rm N}=0.12$, $f_{\rm Si}=0.08$ and $f_{\rm Fe}=0.10$. The results we will obtain turn out to be robust with respect to variations on the values of these input fractions  within a range in agreement with the general picture determined at 10~TeV energies by direct experiments. In Eq.~(\ref{galflux}) each mass component has a rigidity dependent break at energies $\sim Z E_{\rm k}$ towards a steeper spectrum scaling as $E^{-\gamma_2}$, with $\gamma_2\equiv \gamma_1+\Delta\gamma$. The sharpness of this transition is characterized by the parameter $w$ so that, for $w\ll\Delta\gamma$, the transition in which the spectral slope of the component of charge $Z$ changes from $\gamma_1$ to  $\gamma_2$ has a characteristic width of  about $\delta E\simeq (w/\Delta\gamma)ZE_{\rm k}$.  The parametrization in eq.~(\ref{galflux}) is phenomenological, reflecting a change in slope that could be related to a change in the acceleration mechanism in the source or a change in the mechanism of CR escape from the Galaxy. Analogous expressions were for instance adopted previously in \cite{ca02,ho03,tu133,ab18}.

Regarding the extragalactic component, which is in principle expected to become sizable at energies higher than that of the second-knee, it is not our purpose here to model in detail the contributions from the different elements. We will just adopt a single effective component describing the low-energy tail of the extragalactic CRs below one EeV, having a spectrum matching the  power-law with $\gamma\simeq 3.2$ that is observed approaching the ankle. This extragalactic component will be assumed to be exponentially suppressed for decreasing energies, as could be expected in the case in which a magnetic horizon attenuates the contribution from far away sources at low energies \cite{le04,be06,mo13}. This suppression could result from the fact that  at decreasing energies the CRs that diffuse through the random intergalactic fields may take very long times, even longer than the lifetimes of the sources, to reach us from sources that are far away. We then assume the following expression for the extragalactic CR flux
\begin{equation}
\frac{{\rm d}\Phi_{\rm xg}}{{\rm d}E}=\phi_{\rm xg}\left(\frac{E}{\rm EeV}\right)^{-3.2}\frac{1}{{\rm cosh}\left[(E_{\rm T}/E)^\beta\right]}.
\end{equation}
Note that using the hyperbolic cosine rather than just an exponential suppression allows a better match to the asymptotic power-law shape. The energy $E_{\rm T}$ corresponds to a suppression of the extragalactic component by a factor of about 0.65 with respect to the extrapolation of the asymptotic power-law behavior and hence it is close to the energy for which the Galactic and extragalactic contributions to the CR flux are comparable. We will infer the value of the energy at which the transition between a flux dominated by the Galactic component to one dominated by the extragalactic component takes place just from a direct comparison of the two fluxes. The steepness of the suppression of the extragalactic component is determined by the parameter $\beta$.

\begin{figure}[h]
\centering
\includegraphics[scale=.82]{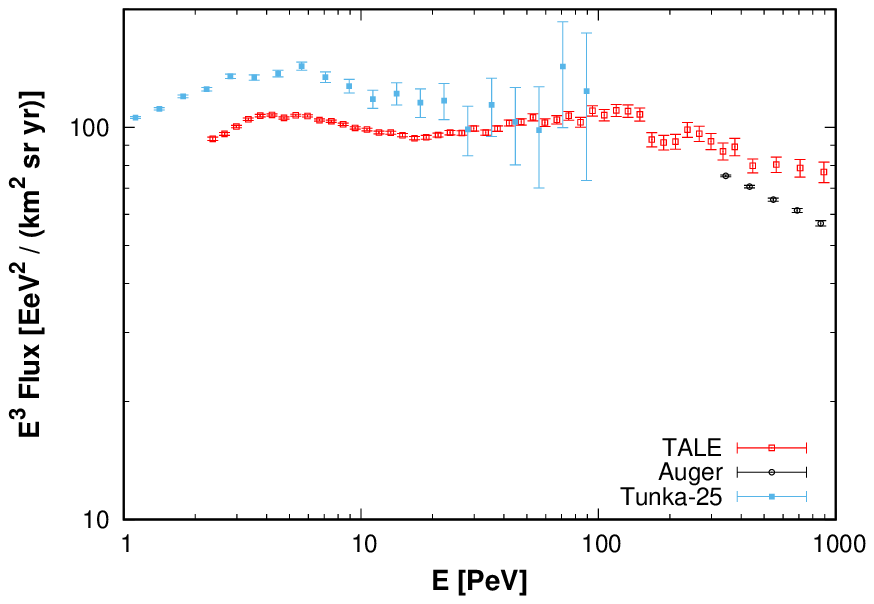}
\includegraphics[scale=.82]{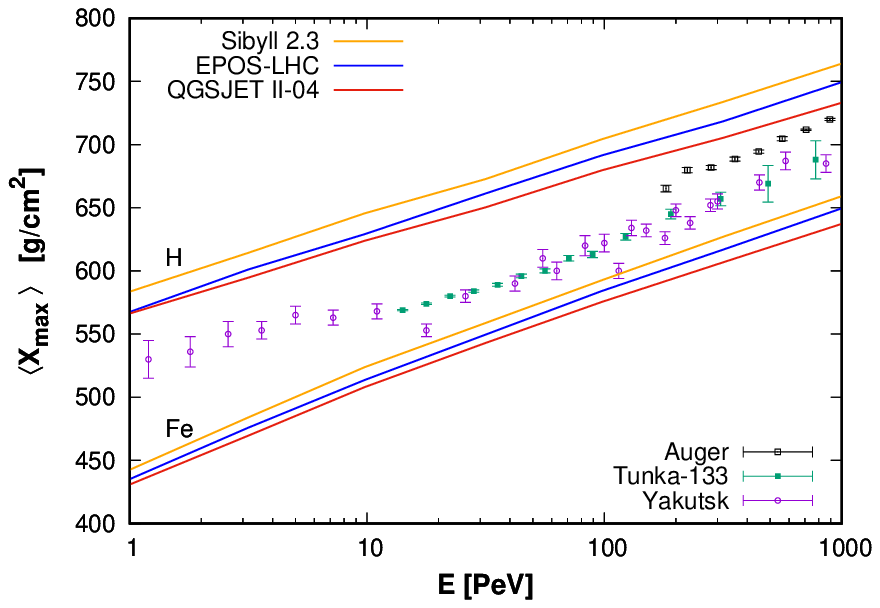}
\caption{Left panel: spectra vs. $E$ measured by Tunka-25 \cite{tu25}, TALE \cite{tale} and Auger \cite{augersp}. The fluxes are multiplied by $E^3$ in order to make them flatter. Right panel: Average $X_{\rm max}$ vs. $E$ measured by Yakutsk \cite{yak}, Tunka-133 \cite{tu133} and Auger \cite{augerxm}. Also shown are the predictions from simulations based on the hadronic models Sibyll 2.3, EPOS-LHC and QGSJet II-04 for pure H and Fe compositions.}
\label{fig:data}
\end{figure}

We will determine the eight parameters $E_{\rm k},\ \gamma_1,\ \gamma_2, \ w$, $E_{\rm T},\ \beta$ and the flux normalizations $\phi_{\rm G}$ and $\phi_{\rm xg}$ through a fit to measurements of the spectrum and of the values of $\langle X_{\rm max}\rangle$ that were mentioned in the Introduction and that are shown in Fig.~\ref{fig:data}. A further issue that is required in order to be able to compare the measured fluxes from different experiments is that there are significant systematic shifts between the energy calibrations performed in each of them, as is apparent from the discrepant flux normalizations that are obtained at a given energy from different datasets. We will conventionally adopt as default energies those measured by the TALE experiment, since it provides the majority of the spectrum data used in this work, and we will then rescale, both in the spectrum and the $X_{\rm max}$ data points,  the Tunka energies by a factor 0.88, those of Auger by a factor 1.07 and those from Yakutsk by a factor 0.625 \cite{be13,de18}.

\section{Composition from the depth of shower maximum}

In order to exploit the composition information contained in the depth of shower-maximum  measurements one needs to adopt a reference hadronic model to compare the observed values of $\langle X_{\rm max}\rangle$ with the expectations for different CR compositions. We show in the right panel of Fig.~\ref{fig:data} the measured $\langle X_{\rm max}\rangle$ as well as the predictions for hydrogen and iron primaries obtained with the models Sibyll 2.3, EPOS-LHC and QGSJet II-04. It is clear that the CR masses that would be required to account for the observations will depend on the model considered, with Sibyll leading to the heaviest inferred masses while QGSJet leading to the lowest values.  It proves convenient to parameterize the model predictions according to
\begin{equation}
X^A_{\rm max}=X^A_0+D^A(E)\,{\rm log}(E/{\rm EeV}),
\label{xmvse}
\end{equation}
with 
\begin{equation}
D^A(E)=D^A_0+D^A_1\,{\rm log}(E/{\rm EeV}).
\label{elongvse}
\end{equation}

\begin{table}[ht]
\centering
\begin{tabular}{c c c c c c c}
\hline\hline
  Model   &  $X_0^{\rm H}$ & $D_0^{\rm H}$ & $D_1^{\rm H}$ &   $X_0^{\rm Fe}$ & $D_0^{\rm Fe}$ & $D_1^{\rm Fe}$  \\
\hline
 Sibyll 2.3 & $762.6 \pm 0.6$ & $58.1 \pm 0.3$ & $-0.5 \pm 0.2$ & $659.3 \pm 0.7$ & $63.2 \pm 0.4$ & $-2.8 \pm 0.2$  \\
 EPOS-LHC & $748.5\pm 0.6$ & $57.4\pm 0.3$&$-0.9\pm 0.2$ & $ 649.9\pm 0.5 $&  $ 63.3\pm 0.3$&  $-2.6 \pm 0.1$ \\
 QGSJet II-04 & $733.7 \pm 0.5$ &  $54.9 \pm 0.2$& $-0.2 \pm 0.1$ & $637.9 \pm 0.7$ & $59.8 \pm 0.4$ &  $-2.9 \pm 0.2$ \\
 \hline
\end{tabular}
\caption{Coefficients of the fits to the $\langle X_{\rm max}\rangle$ vs. log$E$ dependence in eqs.~(\ref{xmvse}) and (\ref{elongvse}) for different hadronic models and for H and Fe CR primaries. All the coefficients are in units of g/cm$^2$.}
\label{tab:hadfit}
\end{table}

We report in Table~\ref{tab:hadfit} the values for the different coefficients obtained by fitting the model predictions in the range between 1~PeV and 100~EeV, both for hydrogen and iron primaries. The agreement between these approximate expressions and the simulation results turn out to be quite good. One can then obtain an estimation of the average of the logarithm of the mass number of the CRs, $\langle {\rm ln}A\rangle$, from the measured average of the depth of shower maximum through the direct interpolation\footnote{The scaling  relation between ln$A$ and $\langle X_{\rm max}\rangle$ is expected from the approximate logarithmic growth of $X_{\rm max}$ with energy and the superposition model for the interaction of CR nuclei. }
\begin{equation}
\langle {\rm ln}A\rangle\simeq {\rm ln}\,56\,\frac{X^{\rm H}_{\rm max}(E)-\langle X_{\rm max}\rangle}{X^{\rm H}_{\rm max}(E)-X^{\rm Fe}_{\rm max}(E)}.
\end{equation}

We will assign to the extragalactic component just an average value of the logarithm of the mass number,  $\langle {\rm ln}A\rangle_{\rm xg}$, to be determined from the fit to the experimental measurements together with all the other parameters appearing in the model. With the available data below 1~EeV it is not possible to obtain more details on the composition of the extragalactic component. 
Note that the composition measurements at few EeV suggest that at these energies the CRs, for which the extragalactic contribution is dominant, should be light. Moreover, if the low energy suppression of the extragalactic component is due to a magnetic horizon effect, one could expect that below the EeV  the extragalactic component should remain quite light because the heavier components would be more strongly suppressed. 

The expected value for $\langle {\rm ln}A\rangle$ for the fitted model is then
\begin{equation}
\langle {\rm ln}A\rangle_{\rm exp}= \frac{1}{{\rm d}\Phi_{\rm tot}/{\rm d}E}\left[\sum_A \frac{{\rm d}\Phi^A_{\rm G}}{{\rm d}E}\,{\rm ln}A+\frac{{\rm d}\Phi_{\rm xg}}{{\rm d}E}\,\langle {\rm ln}A\rangle_{\rm xg}\right],
\end{equation}where $\Phi_{\rm tot}=\Phi_{\rm G}+\Phi_{\rm xg}$.

A delicate issue regarding the $X_{\rm max}$ determinations is that the results from Auger and those from Yakutsk or Tunka are dissimilar in the overlapping energy range (even after accounting for the shift in the energy calibration of each experiment). The differences typically amount to 30 to 40~g/cm$^2$, which would reflect into a change in $\langle {\rm ln}A\rangle$ of order unity. The uncertainties displayed in the figure are the statistical ones, while the systematic ones amount to about 10~g/cm$^2$ in Auger \cite{augerxm}, to about 15--55~g/cm$^2$ in Yakutsk \cite{yak} and 20--40~g/cm$^2$ in Tunka \cite{tu133}. These systematic uncertainties may in principle allow to make the results more compatible.  In the following we will then consider separately the results from Auger and those from Yakutsk and Tunka so as to also understand the possible impact of these systematic effects on the conclusions reached. 

\section{Results}

\begin{figure}
\centering
\includegraphics[width=0.49\textwidth]{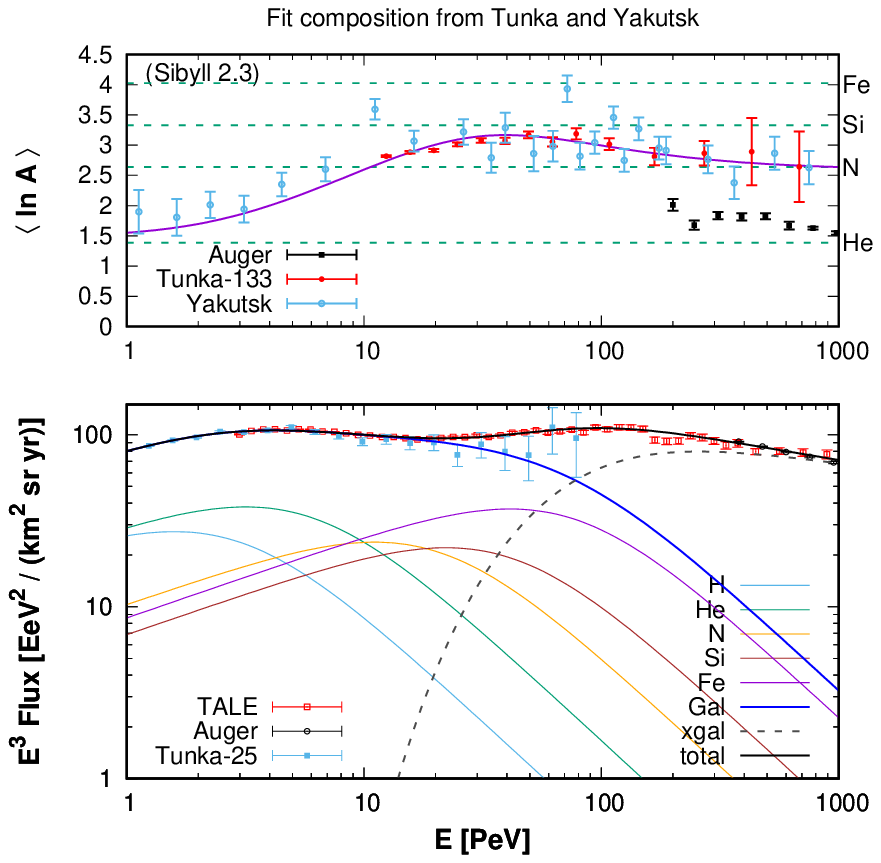}
\includegraphics[width=0.49\textwidth]{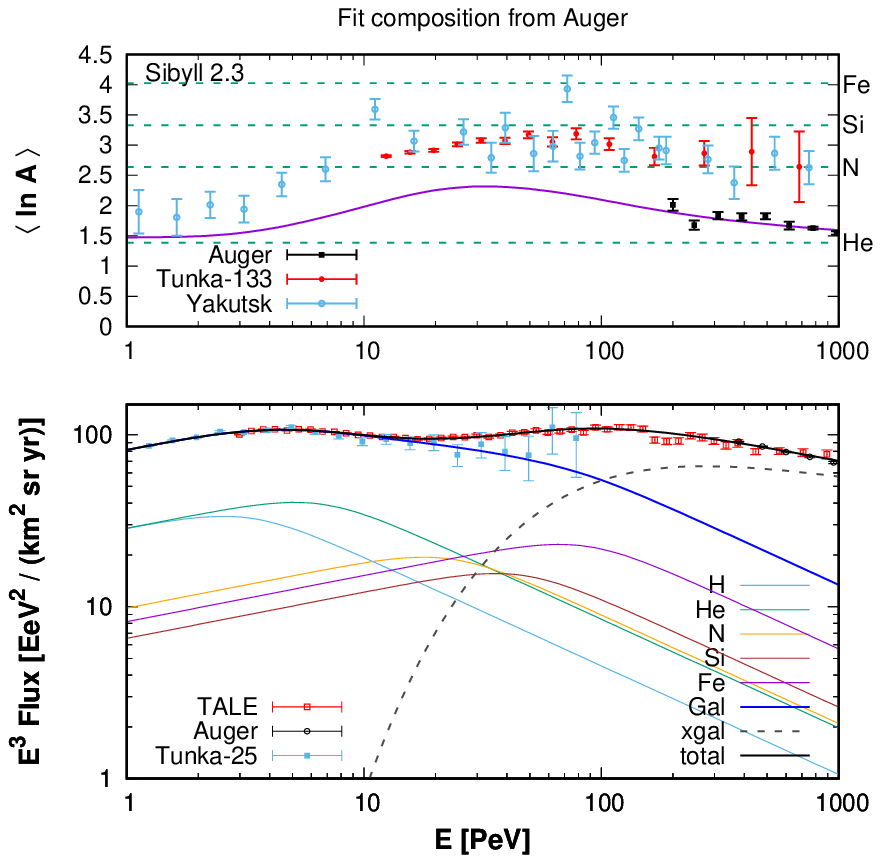}
\includegraphics[width=0.49\textwidth]{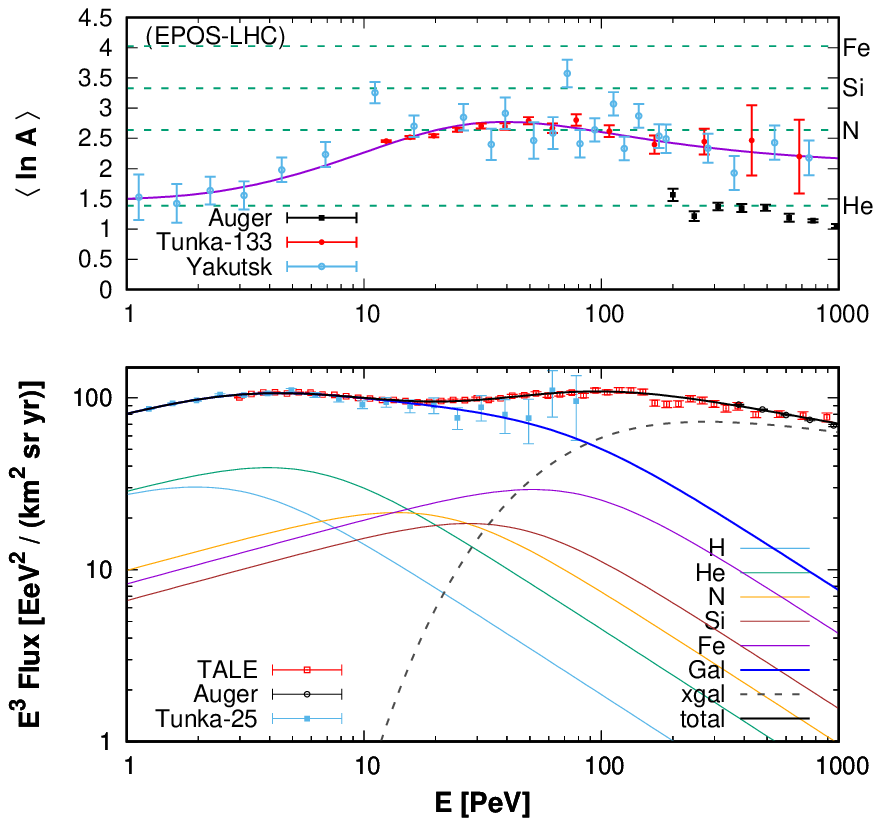}
\includegraphics[width=0.49\textwidth]{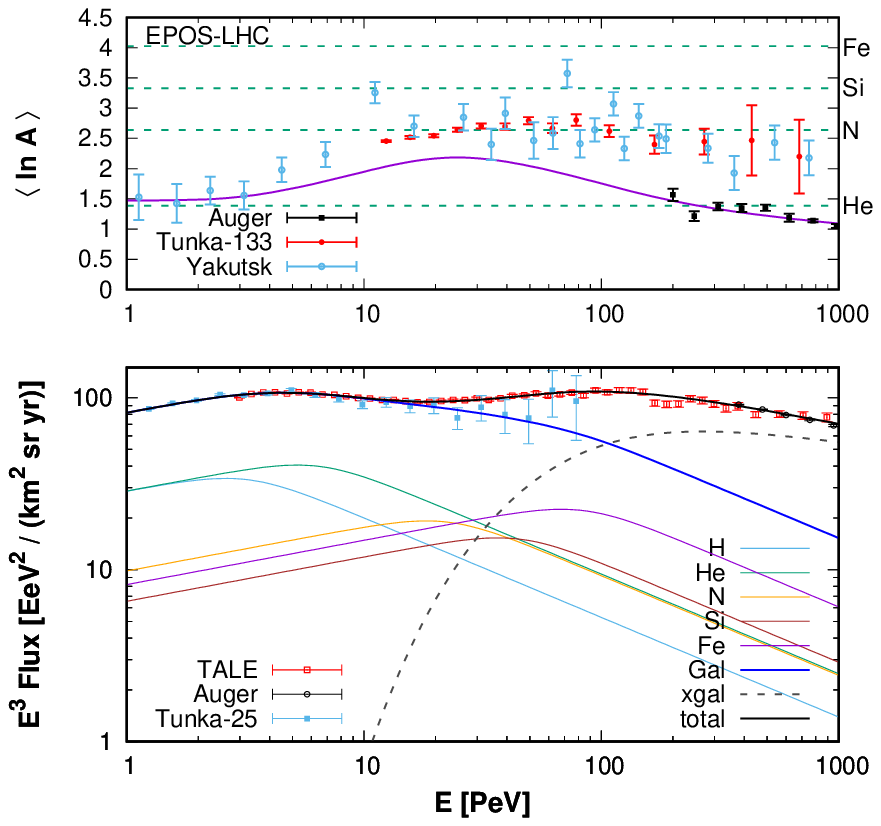}
\includegraphics[width=0.49\textwidth]{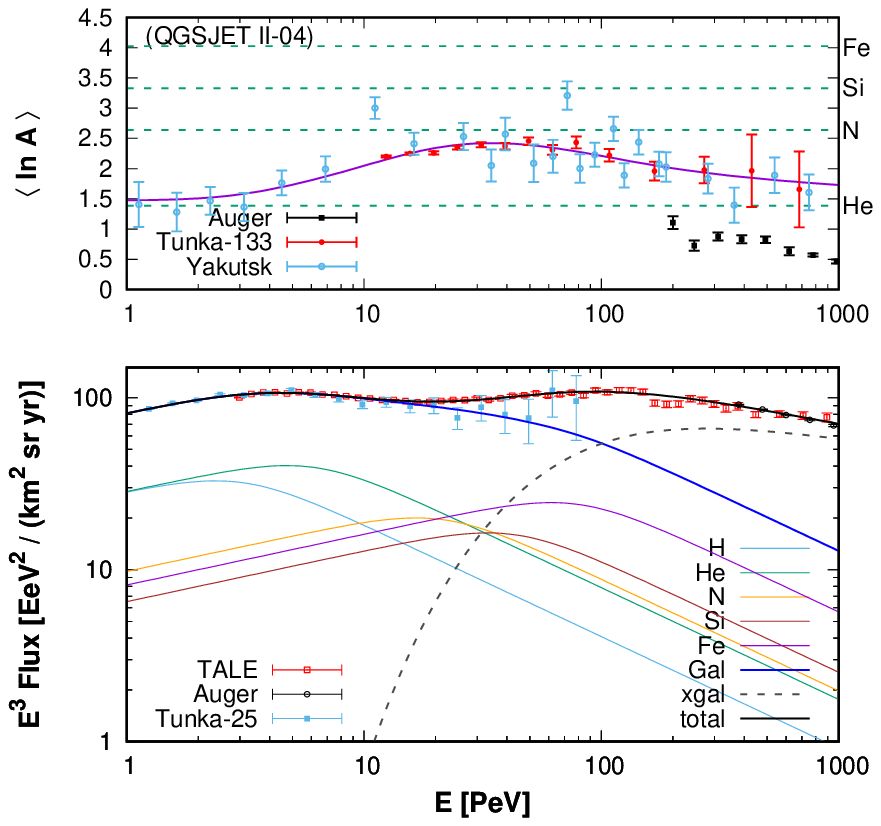}
\includegraphics[width=0.49\textwidth]{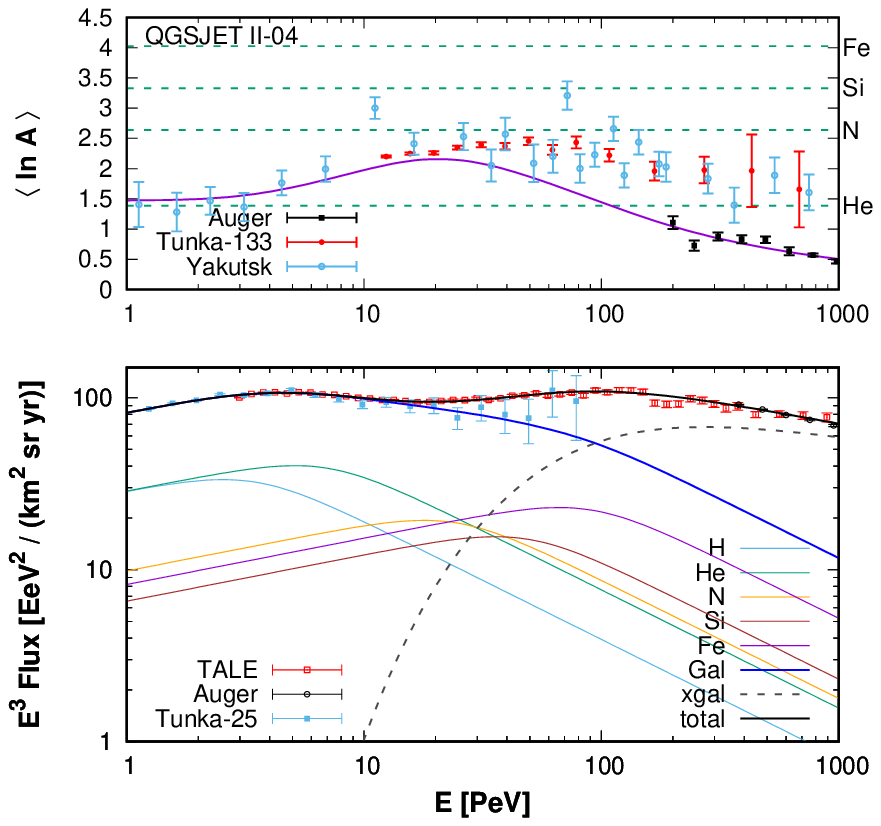}
\caption{Results of the fit to the spectrum and composition data in terms of five Galactic components and an extragalactic one. Plots on the left correspond to the fit to $X_{\rm max}$ data from Tunka and Yakutsk, while those on the right to the fit to Auger data. Plots on the top adopt Sibyll 2.3 hadronic interactions, those in the middle EPOS-LHC ones and those in the bottom are for QGSJet II-04.}\label{fig:sp}
\end{figure}

Using the formalism just introduced we construct a $\chi^2$ function with all the spectrum and $\langle X_{\rm max}\rangle$ measurements in the different energy bins and using MINUIT we determine  the parameters that minimize it. The results obtained for the three hadronic models are shown in Fig.~\ref{fig:sp}. The upper plots correspond to the Sibyll~2.3 hadronic model, the middle ones to EPOS-LHC and the bottom ones QGSJet~II-04. The plots on the left include in the fit the $X_{\rm max}$ data from Yakutsk and Tunka while those on the right include instead the Auger $X_{\rm max}$ data. 
 For each model the top panel shows the results for $\langle  {\rm ln}A\rangle$ while the bottom panel those for the spectrum. The energies of the different experiments have been appropriately rescaled and hence the normalizations of the spectra from different experiments are in agreement. One can see that for all three models a good fit to the data is achieved, with the specific details depending on the hadronic model considered and the $X_{\rm max}$ dataset adopted.  The corresponding parameters for each case are listed in Table~\ref{tab:fit}.\footnote{The uncertainties on the parameters obtained by marginalizing each one with respect to all the remaining ones typically affect the last digit reported. Given the fact that correlations exist between all parameters and that the systematic differences between the values obtained with different hadronic models are larger than the statistical uncertainties, we do not report them for simplicity.}

\begin{table}[ht]
\centering
\begin{tabular}{c | c c c | c c c }
\hline\hline
  & \multicolumn{3}{c|}{$X_{\rm max}$ from Tunka and Yakutsk}& \multicolumn{3}{c}{$X_{\rm max}$ from Auger} \\
& Sibyll 2.3 &    EPOS-LHC &    QGSJet II-04 & Sibyll 2.3 &    EPOS-LHC &    QGSJet II-04 \\
\hline
$E_{\rm k}$  [PeV] & 2.8& 3.0& 3.0 &     3.3& 3.2& 3.4\\
$\gamma_1$ & 2.52 & 2.63& 2.70&       2.73 & 2.74& 2.73 \\
$\gamma_2$ & 4.28 & 3.90& 3.65 &      3.63 & 3.58& 3.69\\
$w$ & 1.0& 0.60& 0.30 &           0.26 & 0.21& 0.30\\
\hline
$\phi_{\rm G}$& 2331& 1096 & 626 &     525 & 486& 528 \\
$\phi_{\rm xg}$ &69 & 64 & 59 &        59 &56 & 61 \\
\hline
$E_{\rm T}$ [PeV] & 115& 121 & 120 &     119 &117 & 120\\
$\beta$ & 0.83& 0.75& 0.73&            0.72 & 0.72& 0.70\\
$\langle {\rm ln}A\rangle_{\rm xg}$ & 2.6& 2.0 & 1.44&      1.27 & 0.60& 0.0\\
\hline
$\chi^2/{\rm dof}$& 2.9& 2.3 & 2.1 &         2.0 & 2.0 & 2.0\\
\hline
\end{tabular}
\caption{Parameters of the  fit to the spectrum and composition data. The fluxes $\phi_{\rm G}$ and $\phi_{\rm xg}$ are in units of  [(km$^2$\,sr\,yr\,EeV)$^{-1}$].  The three columns on the left include the $X_{\rm max}$ data from Tunka and Yakutsk, while those on the right consider the Auger  $X_{\rm max}$ data (the number of degrees of freedom are ${\rm dof}=110$ and 79 respectively). The composition data are interpreted on the basis of different hadronic models in each column.}
\label{tab:fit}
\end{table}

In all the examples the knee is associated with the steepening of the H component that appears at an energy of about 3~PeV. 
Given that at this energy the He component has a flux comparable to the H one, and that this component has a break at around 6~PeV, the steepening of the spectrum takes place actually over a wide energy range.
The low-energy ankle is located at about 20~PeV, where an increasing suppression of the H and He Galactic components is observed, while the heavier ones become increasingly important. However, the observed hardening of the spectrum at this energy actually results from the growing contribution of the extragalactic component, which is anyhow still sub-dominant. In particular, the total Galactic contribution does not show a significant hardening at $E_{\rm la}$.  This possibility has been noted already in \cite{sv14}. The second-knee appears at  $E_{\rm sk}\simeq 26 E_{\rm k}\simeq 80$~PeV and it is associated with the steepening of the Galactic Fe component. This would also be in agreement with the bent in the spectrum of the heavy component that was inferred by the KASCADE-Grande experiment \cite{kg}.

The spectral indices obtained are $\gamma_1\simeq 2.6$--2.7 and $\gamma_2\simeq 3.6$--4.2, and we note that $\gamma_2$ gets essentially fixed by the behavior of the spectrum of the light components between $E_{\rm k}$ and $E_{\rm la}$ rather than by that at $E>E_{\rm sk}$ (similar values of $\gamma_2$ would have been obtained had one restricted the fits to energies below 100~PeV). This fact then determines the steep falloff of the Galactic component beyond the second-knee and implies that the Galactic-extragalactic transition lies just above it, at energies of order 100~PeV.
Note that using Tunka and Yakutsk $X_{\rm max}$ data  the composition becomes quite heavy already at $\sim 20$~PeV, while in the scenarios resulting from the fit to Auger  $X_{\rm max}$ data the increase in $\langle{\rm ln}A\rangle$ is milder. This explains why in the first case the resulting value of $E_{\rm k}$ is slightly smaller (to have an earlier transition), the value of $\gamma_1$ is slightly smaller (to have a faster rise of the heavy components  above the low-energy ankle) and the values of $\gamma_2$ are slightly larger (to have a  faster suppression of the light components beyond their respective knees).

\section{On the impact of a Galactic spectral cutoff}

\begin{figure}
\centering
\includegraphics[width=0.65\textwidth]{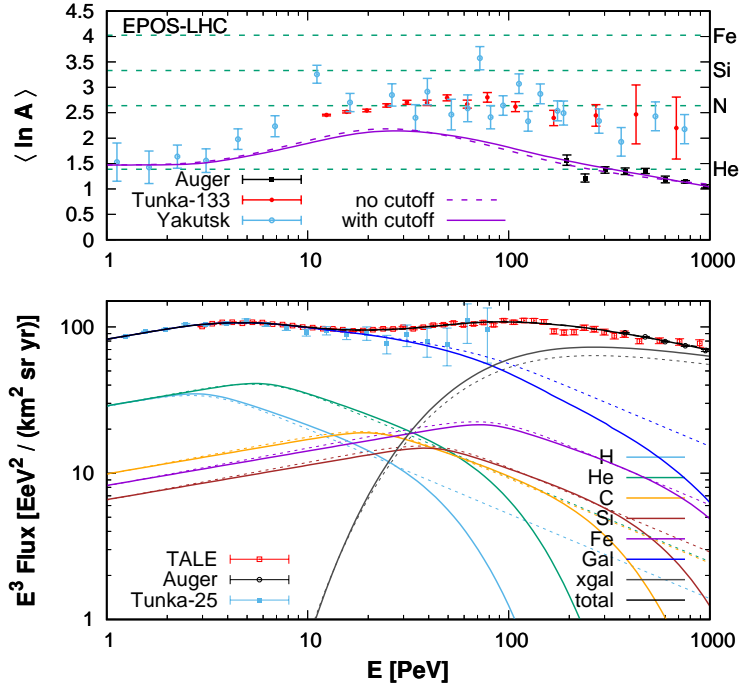}
\caption{Fit to spectrum and composition data including a rigidity dependent cutoff for the Galactic component (solid lines) compared to the results obtained previously without the cutoff (dashed lines). We included in this fit the Auger $X_{\rm max}$ data and adopted the EPOS-LHC hadronic model.}\label{fig:sp_cut}
\end{figure}

The sources of Galactic CRs are expected to reach, sooner or later, a maximum achievable rigidity, limited by the size of the acceleration region and the typical strength of the magnetic fields present in it.  In order to illustrate  the possible impact of these kind of limitations, we explore the effects of introducing an exponential suppression in the flux of the Galactic components beyond a certain maximum rigidity. For definiteness we model the cutoff as
\begin{equation}
\frac{{\rm d}\bar\Phi_G^A}{{\rm d}E}\equiv \frac{{\rm d}\Phi_G^A}{{\rm d}E}\frac{1}{{\rm cosh}(E/ZE_{\rm c})},
\end{equation}
and we determine the cutoff energy $E_{\rm c}$ from the fit, together with all the other parameters discussed before.
The results of performing the fit to the spectrum and composition data (for the case of EPOS-LHC model and considering the Auger $X_{\rm max}$ data) are shown in Fig.~\ref{fig:sp_cut}, where they are also compared to those obtained in the absence of a cutoff (dashed lines in the figure).
The parameters obtained for all the cases (fitting Auger or Tunka and Yakutsk $X_{\rm max}$ data and for the three hadronic models) are reported in Table~\ref{tab:fitcut}. The main differences that appear when the cutoff is introduced are:

\begin{itemize}
\item The fitted values of $E_{\rm c}$ turn out to be in the range 40--60~PeV, i.e. about one decade  above $E_{\rm k}$.

\item The suppression of the Galactic components due to the cutoff gets slightly compensated by  smaller values of $\gamma_2$, and the overall impact on the values of  $\langle{\rm ln}A\rangle$ is small.

\item The suppression of the Galactic component above 100~PeV leads to a slightly larger contribution from the extragalactic CRs above this energy. However, since  the Galactic-extragalactic transition energy is just around this threshold, the transition energy is not significantly affected. Note that for instance the Fe Galactic component only gets suppressed above about 1~EeV.

\item When fitting the Auger $X_{\rm max}$ data the inferred average mass of the extragalactic component gets slightly larger in order to compensate for  the suppression of the contribution from the heavier Galactic component. We also note that the $\chi^2/$dof of the fits slightly improve, having values $\sim 1.8$ for the fits to the Auger $X_{\rm max}$ data. 
 When fitting instead Tunka and Yakutsk  $X_{\rm max}$ data, the inferred value of $\langle{\rm ln}A\rangle_{\rm xg}$ doesn't increase significantly. This can be understood from the fact that in this last case the extragalactic component has an average mass which is already comparable to the total average mass.

\item The fraction of the CR flux of Galactic origin at EeV energies  gets suppressed (by a factor of about two in the example displayed in Fig.~\ref{fig:sp_cut}). This could provide a useful handle to check for the presence of a cutoff by precisely determining the actual contribution of the heavy elements (of Galactic origin) at EeV energies. Note that in the absence of a cutoff the Galactic component contributes about 10--20\% of the total flux at 1~EeV, while in the presence of a cutoff this fraction would be smaller.

\begin{table}[ht]
\centering
\begin{tabular}{c | c c c  | c c c }
\hline\hline
  & \multicolumn{3}{c |}{$X_{\rm max}$ from Tunka and Yakutsk} & \multicolumn{3}{c}{$X_{\rm max}$ from Auger} \\
& Sibyll 2.3 &    EPOS-LHC &    QGSJet II-04 & Sibyll 2.3 &    EPOS-LHC &    QGSJet II-04 \\
\hline
$E_{\rm k}$  [PeV] & 2.8& 2.9& 2.9 &     3.1& 3.1& 3.1\\
$E_{\rm c}$  [PeV] & 61& 53& 41 &     37& 40& 39\\
$\gamma_1$ & 2.52 & 2.63& 2.71&       2.76 & 2.76& 2.76 \\
$\gamma_2$ & 4.26 & 3.85& 3.59 &      3.46 & 3.45& 3.45\\
$w$ & 1.0& 0.56& 0.26 &           0.11 & 0.11& 0.11\\
\hline
$\phi_{\rm G}$& 2334& 1087 & 619 &     425 & 419& 418 \\
$\phi_{\rm xg}$ &70 & 67 & 66 &        65 &64 & 64 \\
\hline
$E_{\rm T}$ [PeV] & 115& 122 & 121 &     117 &117 & 117\\
$\beta$ & 0.82& 0.75& 0.72&            0.73 & 0.74& 0.74\\
$\langle {\rm ln}A\rangle_{\rm xg}$ & 2.6& 2.0 & 1.38&      1.34 & 0.76& 0.11\\
\hline
$\chi^2/{\rm dof}$& 2.9& 2.3 & 2.0 &         1.8 & 1.8 & 1.8\\
\hline
\end{tabular}
\caption{Parameters of the  fit to the spectrum and composition data including a cutoff for the Galactic components. The fluxes $\phi_{\rm G}$ and $\phi_{\rm xg}$ are in units of  [(km$^2$\,sr\,yr\,EeV)$^{-1}$]. The three columns on the left include in the the $X_{\rm max}$ data from Tunka and Yakutsk, while those on the right consider the Auger  $X_{\rm max}$ data, interpreted  on the basis of different hadronic models in each column.}
\label{tab:fitcut}
\end{table}

\end{itemize}

\section{Conclusions}

In recent years an increasingly detailed determination of the CR spectrum and composition was achieved, in particular in the range between 1~PeV and 1~EeV in which the maximum energies of the Galactic accelerators are expected to contribute mostly. The different features observed in the spectrum as well as the changes in the composition can provide the necessary clues to understand the underlying changes in the spectra of each Galactic component, as well as the characteristics of the emerging extragalactic component.

By means of a fit to a selected set of measurements in this energy range within a scenario involving five representative Galactic components, with relative contributions consistent with those measured at 10~TeV energies and  having rigidity dependent broken power-law spectra, we were able to reproduce the different features observed. 
The main insights obtained from this analysis are:

\begin{itemize}
\item At energies below $E_{\rm k}\simeq 3$~PeV the total spectrum has a slope $\gamma_1\simeq 2.6$--2.75 and an average mass with $\langle {\rm ln}A\rangle\simeq 1.5$, consistent with the determinations obtained above TeV energies by direct measurements. 

\item The energy $E_{\rm k}$ represents the steepening of the Galactic H component but, given that the He component provides a comparable contribution to the flux at this energy and that it steepens at an energy of 2$E_{\rm k}\simeq 6$~PeV, the change in the slope of the total spectrum is broad and progressive. Note that a predominant H component at the knee would lead to  a narrower steepening near 3~PeV while a predominant He component would lead to a narrower steepening at about 6~PeV. The results obtained are instead consistent with a comparable amount of both elements being present at the knee (i.e. in similar proportions as measured at 10~TeV), leading then to a broad feature in the spectrum that changes continuously the slope in the range between 3 and 6~PeV.

\item At the low-energy ankle the light components become sub-dominant and the heavy Galactic components get increasingly relevant, leading to a growth in the values of   $\langle {\rm ln}A\rangle$. However,  the hardening of the spectrum at $E_{\rm la}\simeq 20$~PeV is associated with the appearance of the growing extragalactic contribution, since one can see that in Figure~\ref{fig:sp} the total Galactic contribution does not show a hardening at this energy in any of the scenarios considered. 
 The  extragalactic component is relatively light and leads to a decrease in the values of  $\langle {\rm ln}A\rangle$ at energies larger than 30--60~PeV. This is below the energy of the second-knee, at $E_{\rm sk}\simeq 26E_{\rm k}\simeq 80$~PeV, which is associated to the steepening of the Galactic Fe component. The extragalactic contribution becomes dominant above $\sim 100$~PeV. This is in line with the scenarios discussed in \cite{be07b,sv14} and implies also that the ankle hardening at 5~EeV should be a purely extragalactic phenomenon.

\item The inferred value of  $\langle {\rm ln}A\rangle_{\rm xg}$  depends sensitively on the hadronic model adopted as well as on the $X_{\rm max}$ dataset included. Considering the data  from Yakutsk and Tunka one obtains $\langle {\rm ln}A\rangle_{\rm xg}\simeq 2.6$ to 1.4, so that the average extragalactic CR mass ranges from that of N (for Sibyll) to that of He (for QGSJet). 
This is in tension with measurements in the EeV range by Auger \cite{augerxm} and HiRes \cite{hiresxm} that find that for the energies 0.3 to 1~EeV the average composition should not be heavier  than He. Considering Auger composition data one infers that $\langle {\rm ln}A\rangle_{\rm xg}\simeq 1.3$ to 0, consistent with an average mass between that of He (for Sibyll) and H (for QGSJet). This is also more in line with the theoretical  expectations  for the low energy end of the extragalactic component if its suppression is related to  a magnetic horizon effect.

\item The slope of the spectrum of the Galactic components beyond their break, $\gamma_2$, takes values between 3.6 and 4.3, depending on the hadronic models considered. The larger values are obtained for Sibyll, so as to lead to a faster reduction of the light components and hence a more pronounced increase  in the average mass above the knee. Note that the spectral index below the knee, $\gamma_1\simeq 2.6$--2.75, is expected to result from the combined effect of the spectral index at the sources, $\gamma_{\rm s}$, and the energy dependence of the diffusion in the Galaxy, so that for a diffusion coefficient scaling as $D\propto E^\delta$ one has $\gamma_1=\gamma_{\rm s}+\delta$. In particular, for Kolmogorov diffusion one expects $\delta=1/3$ and this would imply that $\gamma_{\rm s}\simeq 2.3$--2.4. The value of the spectral index $\gamma_2$ indicates that beyond the respective knees the steepening with respect to the source spectra is $\Delta\equiv \gamma_2-\gamma_{\rm s}\simeq 1.2$--2. Note that values $\Delta\simeq 1$ are expected if the knee is associated to an enhanced CR escape from the Galaxy due to drift effects related to the regular Galactic magnetic field \cite{pt93,caknee}, while values $\Delta\simeq 2$ would correspond to the transition  to a non-resonant diffusion regime in which the Larmor radius is larger than the maximum scale of the turbulent magnetic field. Hence, settling the issue of the mass composition in this energy range will allow to reach stronger conclusions about the underlying mechanism  explaining the spectral steepening of the Galactic component.

\item We also explored the possible impact of a cutoff in the Galactic components at energies larger than $ZE_{\rm c}$. Fitting the cutoff energy $E_{\rm c}$ we obtained values around 40--60~PeV, so that the Galactic components beyond their knees still follow a power-law with index $\gamma_2$ for almost a decade of energy. It is interesting that the values of $\gamma_2$ get closer to the expectations from drift scenarios (specially when the Auger $X_{\rm max}$ data are considered). The cutoff leads to a strong suppression of the Galactic fraction beyond $26E_{\rm c}$, something that may be tested by detailed determinations of the heavy contribution to the CR flux at EeV energies.  

\item Note that since the Galactic component only provides a contribution of order 10\% at 1~EeV, even if this component is expected to have a significant anisotropy towards the Galactic center and the Galactic plane,  the large scale anisotropies should turn out to be consistent with the bounds obtained  at EeV energies \cite{ab13} as long as the dominant extragalactic component is sufficiently isotropic at these energies. The presence of an exponential suppression of the Galactic component could also further reduce its contribution to the anisotropies.

\end{itemize}

This kind of studies will certainly benefit from additional measurements of the spectrum and of the composition in this energy range, as well as on a better understanding of the systematic effects involved. 

\section*{Acknowledgments}
This work was supported by CONICET (PIP 2015-0369) and ANPCyT (PICT 2016-0660). We thank Tanguy Pierog for providing us the results of the simulations with different hadronic  models and Leonid Kuzmichev for providing us the data from Tunka. We thank the Auger Collaboration for making data available at www.auger.org.

\end{document}